\documentstyle[prl,aps,epsf,multicol,tabularx]{revtex}

\tighten
\begin{document}


\newcommand{\scsc}{\scriptscriptstyle}
\newcommand{\bfx}{\mbox{\boldmath $x$}}
\newcommand{\bfv}{\mbox{\boldmath $v$}}
\newcommand{\bfJ}{\mbox{\boldmath $J$}}
\newcommand{\bfA}{\mbox{\boldmath $A$}} 
\newcommand{\bfsA}{\mbox{\boldmath ${\scriptstyle A}$}} 
\newcommand{\bfE}{\mbox{\boldmath $E$}} 
\newcommand{\bfB}{\mbox{\boldmath $B$}} 
\newcommand{\bfs}{\mbox{\boldmath $s$}} 
\newcommand{\bfr}{\mbox{\boldmath $r$}} 
\newcommand{\bfa}{\mbox{\boldmath $a$}} 
\newcommand{\caS}{{\cal{S}}} 
\newcommand{\friedelphase}{\theta_{\!f}}
\newcommand{\average}[1]{\left\langle \; #1 \; 
   \right\rangle_{n}^{\scsc (t)}} 
\newcommand{\saverage}[1]{\langle \, #1 \, 
   \rangle_{n}^{\scsc (t)}} 
\newcommand{\sumtwo}[2] 
   { \!\!\!\!\!\!\!\! \sum_{\begin{array}{c} {\scriptstyle #1}  
   \\ \left({\scriptstyle #2}\right) \end{array} } \!\!\!\! }  

\newcommand{\sect}[1]
   {\vspace{0.3cm} $\langle$ {\it #1} $\rangle$ }  

\makeatletter 
   \def\@biblabel#1{(#1)} 
   \def\refname{\vspace{0.5cm} 
   \noindent {\it References}  
   \vspace{-0.2cm}} 
   \makeatother

\baselineskip 0.39cm

\draft

\title{Charge Current Density from the Scattering Matrix}

\author{Tooru Taniguchi}

\address{D\'epartement de Physique Th\'eorique, 
   Universit\'e de Gen\`eve, \\ CH-1211, Gen\`eve 4, Switzerland}

\date{\today}

\maketitle

\begin{abstract}

   A method to derive the charge current density and its quantum
mechanical correlation from the scattering matrix is discussed
for quantum scattering systems described by a time-dependent
Hamiltonian operator. 
   The current density and charge density are expressed with the 
help of functional derivatives with respect to the vector 
potential and the electric potential. 
   A condition imposed by the requirement that these local 
quantities are gauge invariant is considered. 
   Our formulas lead to a direct relation between the local 
density of states and the total current density at a given energy.
   To illustrate the results we consider, as an example, a 
chiral ladder model. 

\end{abstract}

\vspace{0.5cm}

\begin{multicols}{2}

\narrowtext


\sect{Introduction}
   The scattering matrix gives an important starting point in 
descriptions of quantum transport phenomena. 
   It connects the incoming current amplitudes to the outgoing 
current amplitudes, and is calculated from the Hamiltonian operator 
by the M{\o}ller operator or the Green function method \cite{joa75}. 
   We can obtain information about {\em global} characteristics 
in systems from the scattering matrix directly. 
   For example, the Landauer formula gives a method to calculate 
the conductance from transmission amplitudes as components of  
the scattering matrix \cite{lan70,eco81,fis81,but85,but86,tan98}. 
   The Friedel sum rule connects the density of states to the 
scattering matrix \cite{fri52,langar61}, so that 
it is in principle possible to calculate any equilibrium statistical 
mechanical quantity from the scattering matrix \cite{das69}.
   We can also derive an expression of the persistent current 
in open conductors caused by a magnetic flux \cite{akk91}, 
or the shot noise \cite{les89,but92} from the scattering 
matrix. 

   It should be emphasized that the scattering matrix even gives 
information about {\em local} characteristics in systems. 
   We consider one-particle and time-dependent Hamiltonian systems.  
   First, the charge current density $\bfJ_{n}(\bfx,t)$ at position 
$\bfx$ and time $t$ generated by the particle incident in a state 
with the quantum number $n$ is connected to the scattering 
matrix $S=(S_{nn'})$ as  

\begin{eqnarray}
   \bfJ_{n}(\bfx,t) = \frac{c}{2\pi i} \sum_{n'} S_{n'n}^{*} 
   \frac{\delta S_{n'n}}{\delta \bfA(\bfx,t)}
\label{Curre}\end{eqnarray} 

\noindent where $\bfA(\bfx,t)$ is the vector potential and $c$ 
is the velocity of light.  
   In this Letter we call this quantity the "injected current 
density". 
   Eq. (\ref{Curre}) is the main result of this Letter. 
   Second, the probability density $\rho_{n}(\bfx,t)$ of the 
state caused by the incident particle with the quantum number 
$n$ is given by 

\begin{eqnarray}
   \rho_{n}(\bfx,t) = - \frac{1}{2\pi i} \sum_{n'} S_{n'n}^{*} 
   \frac{\delta S_{n'n}}{\delta U(\bfx,t)} 
\label{Proba}\end{eqnarray} 

\noindent with the potential $U(\bfx,t)$. 
   The probability density $\rho_{n}(\bfx,t)$ also has been called 
the "injectivity" and can be regarded as the density of states with a 
preselection of the incident channel. 
   Eq. (\ref{Proba}) in the time-independent Hamiltonian case has 
been used in treatments of the electron-electron interaction using 
a self-consistent potential \cite{but93,but98,chr97,gra99}.  
   In this Letter we give a new derivation of this  
formula and generalize it to the time-dependent Hamiltonian 
case.  

   Our technique to derive Eqs. (\ref{Curre}) and (\ref{Proba})
also can be used to arrive at quantum mechanical correlation 
functions of local quantities from the scattering matrix. 
   For example we show that 
the quantum mechanical correlation 
$C_{n}^{\scsc (\mu\nu)}(\bfx,\bfx';t)$ 
between the $\mu$-th charge current density component  at position 
$\bfx$ and the $\nu$-th charge current density component at position 
$\bfx'$ at the same time $t$ is given by 

\begin{eqnarray}
   && C_{n}^{\scsc (\mu\nu)}(\bfx,\bfx';t) 
   = \frac{c^{2}\hbar}{2\pi} \sum_{n'} \mbox{Re}\left\{ 
   \frac{\delta S_{n'n}^{*}}{\delta A^{\scsc (\mu)}(\bfx,t)} 
   \frac{\delta S_{n'n}}{\delta A^{\scsc (\nu)}(\bfx',t)}\right\} 
   \nonumber \\ 
   &&
\label{CurreCorre} \end{eqnarray}

\noindent where $A^{\scsc (\mu)}(\bfx,t)$ is the $\mu$-th component 
of the vector potential.  

   Our formulas include the vector potential and the potential 
explicitly, so we have to discuss the gauge invariance. 
   We show Eqs. (\ref{Curre} - \ref{Proba}) 
and (\ref{CurreCorre}) to be gauge invariant, and 
derive conditions which should be satisfied by locally gauge invariant  
quantities.

   In the time-dependent Hamiltonian system an energy of an outgoing 
particle can be different from the energy of the corresponding 
incoming particle, and the scattering matrix element $S_{nn'}$ 
describes a transition to such a different energy.  
   On the other hand, if we consider only time-independent 
Hamiltonian cases, the problem becomes simpler, because 
the scattering matrix is decomposed into the scattering matrices 
restricted to the energy shells. 
   Using this feature we obtain a relation between the local 
density of states and the total current density defined 
by the sum of the injected current density $\bfJ_{n}(\bfx,t)$ 
with respect to the suffix $n$ satisfying the condition $E_{n}=E$ 
at energy $E$. 

   As a simple example we investigate a ladder model with 
a directionality, namely a chiral ladder model, termed a "quantum 
rail road" in Ref. \cite{bar93}. 
   We verify the formula (\ref{Curre}) for this model   
calculating separately the current density and 
the scattering matrix.


\sect{Injected current density and injectivity} 
   We start from the time-dependent Hamiltonian operator 
$\hat{H}(t)$ in the quantum scattering system. 
   The dynamics of the system is described by the Schr\"odinger 
equation using this Hamiltonian operator. 
   We decompose the total Hamiltonian operator $\hat{H}(t)$ into 
the asymptotic Hamiltonian operator $\hat{H}_{0}$ and the scattering 
operator $\hat{H}_{1}(t)$; $\hat{H}(t)=\hat{H}_{0}+\hat{H}_{1}(t)$. 
   The operator $\hat{H}_{0}$ is chosen to be the Hamiltonian 
operator which describes incoming and outgoing particles in the 
asymptotic regions.  
   We assume that $\hat{H}_{0}$ is a time-independent 
operator determined uniquely. 
   Besides, the system described by the Hamiltonian operator 
$\hat{H}(t)$ or $\hat{H}_{0}$ is assumed to have no bound state.  
   Below we use the coordinate representation for any operator 
and take $\bfx$ as the coordinate of particles. 
   Under these conditions the scattering matrix element $S_{nn'}$ 
is given by  
   
\begin{eqnarray}
   S_{nn'} = 
   \lim_{t_{2}\rightarrow +\infty}\lim_{t_{1}\rightarrow -\infty} 
   \int d\bfx\; \Phi_{n}(\bfx)^{*} \hat{U}(t_{2},t_{1}) 
   \Phi_{n'}(\bfx)   
\label{ScattMatri}\end{eqnarray} 
   
\noindent where $\hat{U}(t_{2},t_{1})$ is the time evolution 
operator $\exp\{i\hat{H}_{0}t_{2}/\hbar\}$$\cdot\vec{T} 
\exp\{-i\int_{t_{1}}^{t_{2}}dt\hat{H}(t) /\hbar\}$  
$\cdot \exp\{-i$ $\hat{H}_{0}t_{1}/\hbar\}$ in 
the interaction picture with $\vec{T}$ being the positive 
time-ordering operator, and $\Phi_{n}(\bfx)$ is the eigenstate 
of the operator $\hat{H}_{0}$ corresponding to the energy 
eigenvalue $E_{n}$. 
   Here the limits $t_{1} \rightarrow -\infty$ and $t_{2} 
\rightarrow +\infty$ are defined by $\lim_{t\rightarrow\pm\infty} 
X(t) \equiv \lim_{\epsilon\rightarrow +0} (\pm\epsilon)
\int_{0}^{\pm\infty}dt' e^{\mp\epsilon t'}$$X(t')$ for any 
function $X(t)$ of $t$ \cite{joa75}.
   The set $\{\Phi_{n}(\bfx)\}_{n}$ of the eigenstates of 
the operator $\hat{H}_{0}$ is chosen to satisfy the 
orthonormality condition and the completeness relation.  
   The scattering matrix $S=(S_{nn'})$ is shown to be an 
unitary matrix; $S^{\dagger}S = SS^{\dagger} = I$. 

   For simplicity we consider the one-particle system. 
   The total Hamiltonian operator $\hat{H}(t)$ and the operator 
$\hat{H}_{0}$ are represented as 
$( -i\hbar\partial/\partial \bfx - q \bfA(\bfx,t)/c )^2 /(2m) 
+ U(\bfx,t)+U_{0}(\bfx)$ and $-(\hbar^{2}/(2m)) 
\partial^{2}/\partial \bfx^{2} + U_{0}(\bfx)$, respectively, 
where $m$ is the mass of the particle, $q$ is the charge of the 
particle,  $\bfA(\bfx,t)$ is the vector potential, and 
$U(\bfx,t)$ is the external potential (plus the induced 
potential by the interaction of particles), and $U_{0}(\bfx)$ is
the confinement potential. 
   We consider a functional derivative $\delta^{\scsc (t)} 
\hat{S}_{nn'}$ of the scattering matrix element with respect to 
the vector potential $\bfA(\bfx,t)$ or the potential $U(\bfx,t)$ 
at time $t$, using the notation $\delta^{\scsc (t)} = 
\delta/\delta\bfA(\bfx,t)$ or $\delta/\delta U(\bfx,t)$. 
   Using the expression (\ref{ScattMatri}) of the scattering 
matrix elements we obtain a general relation 

\begin{eqnarray}
   \frac{1}{2\pi i} \sum_{n'} S_{n'n}^{*} (\delta^{\scsc (t)}S_{n'n})
   = - \average{\int_{-\infty}^{\infty}dt'\; 
   \delta^{\scsc (t)} \hat{H}_{1}(t')}.
\label{Sumrule1}\end{eqnarray} 

\noindent Here the notation $\saverage{\cdots}$ 
means the expectation value taken with the scattering state 
of the quantum number 
$n$, namely $\saverage{\hat{X}} \equiv \int d\bfx  
\Psi_{n}(\bfx,t)^{*} \hat{X} \Psi_{n}(\bfx,t)$ 
$/(2\pi\hbar)$ for any operator $\hat{X}$, where $\Psi_{n}(\bfx,t)$ 
is a solution of the Schr\"odinger equation using the total 
Hamiltonian operator $\hat{H}(t)$ and is defined by 
$\Psi_{n}(\bfx,t) \equiv  \lim_{t_{1} \rightarrow -\infty}  
   \vec{T}\exp\{-i\int_{t_{1}}^{t}dt'\;\hat{H}(t')/\hbar\} 
   \cdot \exp\{-iE_{n}t_{1}/\hbar\} \Phi_{n}(\bfx)$.
The derivation of Eq. (\ref{Sumrule1}) is given by using  
the completeness relation of the set $\{\Phi_{n}(\bfx)\}_{n}$, the 
property $\hat{U}(t_{2},t_{1})=\hat{U}(t_{2},t)\hat{U}(t,t_{1})$ 
of the time evolution operator and the relation 
$\delta^{\scsc (t)}\hat{U}(t_{2},t_{1}) = \hat{U}(t_{2},t) \cdot 
\exp{\{i\hat{H}_{0}t/\hbar\}} \cdot \{(-i/\hbar)\; 
\int_{-\infty}^{\infty}dt' \delta^{\scsc (t)} \hat{H}_{1}(t')\} 
\cdot \exp{\{-i\hat{H}_{0}t/\hbar\}} \cdot \hat{U}(t,t_{1})$ in 
$t_{1}<t<t_{2}$. 
   Eq. (\ref{Sumrule1}) is a key result of this Letter. 
   We introduce the injected current density $\bfJ_{n}(\bfx,t)$ and 
the injectivity $\rho_{n}(\bfx,t)$ as  

\begin{eqnarray}
   \bfJ_{n}(\bfx,t) 
   \equiv \average{\hat{\bfJ}(\bfx,t)}, 
   \;\;\;\;\;\;    
   \rho_{n}(\bfx,t) 
   \equiv \average{\hat{\rho}(\bfx)}
\label{ProbaCurre}\end{eqnarray} 

\noindent using the charge current density operator 
$\hat{\bfJ}(\bfr,t) \equiv q \hat{\bfv}(t) \star 
\delta(\bfx-\bfr)$ with $\hat{\bfv}(t)$ being the velocity operator 
$( -i\hbar\partial/\partial \bfx - q\bfA(\bfx,t)/c )/m$ and the 
multiplication $\star$ being the symmetrized product, and using the 
probability density operator $\hat{\rho}(\bfr) \equiv \delta 
(\bfx-\bfr)$. 
   Eqs. (\ref{Curre}) and (\ref{Proba}) are derived from 
Eq. (\ref{Sumrule1}), using the relations 
$\int_{-\infty}^{\infty}dt' \delta\hat{H}_{1}(t') 
/\delta\bfA(\bfx,t) = -\hat{\bfJ}(\bfx,t)/c$ and 
$\int_{-\infty}^{\infty}dt' \delta\hat{H}_{1}(t') 
/\delta U(\bfx,t) = \hat{\rho}(\bfx)$. 


\sect{Current density correlation} 
   We consider functional derivatives 
$\delta_{j}^{\scsc (t)}S_{nn'}$, $j=1,2$ of the scattering 
matrix element with respect to the vector potential 
$\bfA(\bfx,t)$ or the potential $U(\bfx,t)$ at time $t$. 
   Using Eq. (\ref{ScattMatri}) we obtain another general relation 

\begin{eqnarray}   
   && \hspace{0cm} 
      \frac{\hbar}{2\pi} 
      \sum_{n'} \left( \delta_{1}^{\scsc (t)} S_{n'n}^{*} \right) 
      \left( \delta_{2}^{\scsc (t)} S_{n'n} \right) 
      \nonumber \\ 
   && =   \average{ 
      \left( \int_{-\infty}^{\infty}dt' 
        \delta_{1}^{\scsc (t)} \hat{H}_{1}(t') \right) 
      \left( \int_{-\infty}^{\infty}dt'' 
        \delta_{2}^{\scsc (t)} \hat{H}_{1}(t'') \right) 
      }. 
\label{Sumrule2}\end{eqnarray} 

\noindent Using  Eq. (\ref{Sumrule2}) we obtain Eq. (\ref{CurreCorre}). 
   Here the quantum mechanical current density correlation 
$ C_{n}^{\scsc (\mu\nu)}(\bfx,\bfx';t) $ is defined by   

\begin{eqnarray}
   C_{n}^{\scsc (\mu\nu)}(\bfx,\bfx';t) \equiv \average{
   \hat{J}^{\scsc (\mu)}(\bfx,t) 
   \star \hat{J}^{\scsc (\nu)}(\bfx',t) }
\label{CurreDensiCorre}\end{eqnarray}

\noindent where $\hat{J}^{\scsc (\mu)}(\bfx,t)$ is the $\mu$-th 
component of the charge current density operator. 
   It should be noted that the average $\saverage{\cdots}$ taken in 
the correlation (\ref{CurreDensiCorre}) includes only the quantum 
mechanical average, but does not include the thermo-statistical 
average.

   In similar ways we can obtain other quantum mechanical 
correlation functions $\saverage{\hat{J}^{\scsc (\nu)}(\bfx,t) \star  
\hat{\rho}(\bfx')}$ and $\saverage{\hat{\rho}(\bfx) \star 
\hat{\rho}(\bfx')}$ from the scattering matrix. 


\sect{Gauge invariance} 
   The gauge transformation of the electric potential $\phi(\bfx,t)$ 
and the vector potential $\bfA(\bfx,t)$ is represented as 
$\phi(\bfx,t)\rightarrow\phi(\bfx,t)+\Delta \phi(\bfx,t)$ 
and $\bfA(\bfx,t)\rightarrow\bfA(\bfx,t)+\Delta \bfA(\bfx,t)$ with   
$\Delta \phi(\bfx,t) \equiv -(1/c) \partial \varphi(\bfx,t) / 
\partial t$ and $\Delta \bfA(\bfx,t) \equiv \partial \varphi(\bfx,t) 
/\partial \bfx$ using a function $\varphi(\bfx,t)$ of $\bfx$. 
   The electric and magnetic fields are invariant under 
the gauge transformation.    
   Using that the Schr\"odinger equation is gauge invariant, 
we can show that the $\mu$-th component 
$J_{n}^{\scsc (\mu)}(\bfx,t)$ of the injected current  
density is also gauge invariant, meaning that the formula 
(\ref{Curre}) is gauge invariant. 

   The fact that the injected current density is gauge invariant 
is also represented as 
   
\begin{eqnarray}
   && \int dt'\; \int d\bfx'\; \left\{ 
   \frac{\delta J_{n}^{\scsc (\mu)}(\bfx,t)}{\delta\phi(\bfx',t')} 
   \; \Delta \phi(\bfx',t') \right. \nonumber \\ 
   &&\hspace{2.7cm} 
   + \left.\frac{\delta J_{n}^{\scsc (\mu)}(\bfx,t)}{\delta\bfA(\bfx',t')} 
   \cdot \Delta \bfA(\bfx',t') \right\} = 0. 
\label{GaugeInvar1} \end{eqnarray} 
   
\noindent A change $\delta\phi(\bfx,t)$ of the electric potential 
$\phi(\bfx,t)$ modifies the potential $U(\bfx,t)$ as 
$\delta U(\bfx,t) = q\delta\phi(\bfx,t)$. 
   Moreover Eq. (\ref{GaugeInvar1}) should be satisfied for any 
function $\varphi(\bfx,t)$ which is zero at the boundary 
of the integral of the left-hand side of Eq. (\ref{GaugeInvar1}). 
   Using these facts and the partial integrals in Eq. 
(\ref{GaugeInvar1}) we obtain

\begin{eqnarray}
   \frac{q}{c} \frac{\partial}{\partial t'} 
   \frac{\delta J_{n}^{\scsc (\mu)}(\bfx,t)}{\delta U(\bfx',t')} 
   + \frac{\partial}{\partial \bfx'} \cdot 
   \frac{\delta J_{n}^{\scsc (\mu)}(\bfx,t)}{\delta\bfA(\bfx',t')} 
   =0.  
\label{GaugeInvar3}\end{eqnarray}
 
\noindent Eq. (\ref{GaugeInvar3}) represents a condition for the 
injected current density imposed by its gauge invariance. 
   Similarly, the formulas (\ref{Proba}) and (\ref{CurreCorre}) 
are gauge invariant, and the injectivity $\rho_{n}(\bfx,t)$ satisfies 
an equation similar to Eq. (\ref{GaugeInvar3}).


\sect{Total current density and local density of states in 
the time-independent Hamiltonian system} 
   Now, we consider the time-independent Hamiltonian case. 
   In this case the injected current density and the injectivity 
are independent of time $t$. 
   Moreover the state $\Psi_{n}(\bfx,t)$ used to define the average 
$\saverage{\cdots}$ can be replaced by the 
state $\bar{\Psi}_{n}(\bfx) \equiv \{1+\hat{G}(E_{n})\hat{H}_{1} \} 
\Phi_{n}(\bfx)$ using the Green operator $\hat{G}(E) \equiv 
\lim_{\epsilon\rightarrow +0}(E-\hat{H} + i\epsilon)^{-1}$. 
   The state $\bar{\Psi}_{n}(\bfx)$ is the eigenstate of the 
total Hamiltonian operator $\hat{H}$ corresponding to the 
eigenvalue $E_{n}$ and satisfies the Lippmann-Schwinger equation. 
   One of the important features in the time-independent 
Hamiltonian system is that the scattering matrix element $S_{nn'}$ 
takes a non-zero value only in the case of $E_{n} = E_{n'}$. 
   Therefore, in the formulas (\ref{Curre}) and
(\ref{Proba}) we can replace the full scattering matrix element 
$S_{n'n}$ with the matrix element $\caS_{l'l}(E_{n})$ of the 
scattering matrix $\caS(E_{n})$ restricted to the energy shell 
of energy $E_{n}$, and exchange the sum of the quantum number $n'$ 
with the sum of the channel number $l'$.
  
  It is important to note a difference between the time-dependent 
and time-independent Hamiltonian cases in our formulas. 
   In the time-dependent Hamiltonian case the functional derivative 
$\delta S_{nn'}/ \delta \bfA(\bfx,t)$ is given by 
$\delta S_{nn'} = \int d\bfx \int dt $ 
$(\delta S_{nn'}/ \delta \bfA(\bfx,t)) \delta \bfA(\bfx,t)$.  
   On the other hand, in the time-independent Hamiltonian case 
the functional derivative $\delta \caS_{ll'}(E)/ \delta\bfA(\bfx)$ 
is given by $\delta \caS_{ll'}(E) = \int d\bfx (\delta 
\caS_{ll'}(E)/ 
\delta\bfA(\bfx)) \delta \bfA(\bfx)$, which does not include the 
time integral like in the time-dependent Hamiltonian case.   
   This difference appears in the current density correlation formula 
(\ref{CurreCorre}), where a simple exchange of the scattering 
matrix element $S_{nn'}$ and the vector potential $\bfA(\bfx,t)$ 
with the on-shell scattering matrix element $\caS_{ll'}(E)$ and the 
vector potential $\bfA(\bfx)$, respectively, is not allowed 
to obtain its time-independent Hamiltonian version. 
   However we can obtain the time-independent 
Hamiltonian versions of the injected current density formula
(\ref{Curre}) and the injectivity formula (\ref{Proba}) by  
such a simple exchange. 
   This technical point is better considered in a separate paper.
    
   We proceed to consider the total current density 
$\bfJ(E,\bfx)$ and the local density of states 
$\rho(E,\bfx)$ defined by  
  
\begin{eqnarray}
   \bfJ(E,\bfx) \equiv \sumtwo{n}{E_{n}=E} \bfJ_{n}(\bfx),  
   \;\;\;\;\;\; 
   \rho(E,\bfx) \equiv \sumtwo{n}{E_{n}=E} \rho_{n}(\bfx).
\label{Total}\end{eqnarray} 

\noindent Eqs. (\ref{Curre}) and (\ref{Proba}) lead to 
   
\begin{eqnarray}
   \bfJ(E,\bfx) = \frac{c}{\pi} 
   \frac{\delta \friedelphase(E)}{\delta \bfA(\bfx)}, \;\;\;\;\;\; 
   \rho(E,\bfx) = - \frac{1}{\pi} 
   \frac{\delta \friedelphase(E)}{\delta U(\bfx)} 
\label{Densi}\end{eqnarray} 
 
\noindent where $\friedelphase(E)$ is called "Friedel phase"
\cite{tan99} and is defined by $\friedelphase(E) \equiv (1/(2i))
\ln \mbox{Det} \{\caS(E)\}$. 
   It may be noted that using the first equation in 
Eq. (\ref{Densi}) the relation $\bfJ(E,\bfx) = 
-\bfJ(E,\bfx)|_{\bfsA \rightarrow -\bfsA}$ is obtained under 
the condition $S_{nn'}=S_{n'n}|_{\bfsA \rightarrow -\bfsA}$.
   The second formula in Eq. (\ref{Densi}) have already 
been used in some works \cite{but93,but98,chr97,gra99,but94}. 
   It follows from Eq. (\ref{Densi}) that

\begin{eqnarray}
    \frac{\delta \bfJ(E,\bfx')}{\delta U(\bfx)} 
    = -c \frac{\delta \rho(E,\bfx)}{\delta \bfA(\bfx')}.
\label{Relat}\end{eqnarray} 

\noindent This equation shows a connection of 
the current density to the local density of states.

  The total current density $\bfJ(E,\bfx)$ takes a non-zero 
value in some important phenomena, such as edge currents in the  
quantum hall effect, persistent currents caused by a magnetic 
flux and so on.
   Earlier work \cite{akk91} obtained the total persistent 
current from the flux derivative of the scattering matrix for 
a ring connected to a lead \cite{but85a}. 
   For a purely one-dimensional ring the functional derivative 
with respect to the vector potential can simply be replaced by 
the derivative with respect to the flux. 
   In such a special case the first formula in Eq. (\ref{Densi}) 
leads to the same expression as the earlier work.

\begin{figure}[t]
   \epsfxsize 7cm 
   \centerline{\epsffile{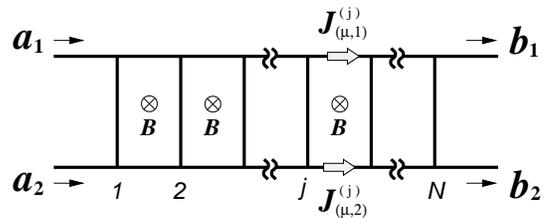}}
   \vspace{0cm}
   \caption{Chiral Ladder Model.} 
   \vspace{-0.2cm}
   \label{ladde} 
\end{figure}

\sect{Example} 
   As a specific example we consider a chiral ladder model threaded 
by a weak magnetic field. 
   In this model particles move on the both legs of the ladder 
only in one direction (See Fig. \ref{ladde}). 
   This may be regarded as a model of two edge channels at one 
edge of a quantum Hall bar with impurities, by which electrons 
transfer from an edge channel to another channel.  

   We assume that the ladder has one-dimensional legs each 
of which has one channel. 
   We introduce the scattering matrix $T^{\scsc (j)}$ which connects 
the current amplitude to the left of the $j$-th rung to the 
current amplitude to the left of the $j\!+\!1$-th rung. 
   The dependence of the scattering matrix $T^{\scsc (j)}$ on
the vector potential in the legs is 

\begin{eqnarray} 
   T^{\scsc (j)} =    
   \left(\begin{array}{cc}  
      t_{11}^{\scsc (j)}\exp\{i\phi_{1}^{\scsc (j)}/\phi_{0}\} &
      t_{12}^{\scsc (j)}\exp\{i\phi_{1}^{\scsc (j)}/\phi_{0}\} \\ 
      t_{21}^{\scsc (j)}\exp\{i\phi_{2}^{\scsc (j)}/\phi_{0}\} &
      t_{22}^{\scsc (j)}\exp\{i\phi_{2}^{\scsc (j)}/\phi_{0}\}  
   \end{array} \right) 
\label{TransMatri}\end{eqnarray} 

\noindent where $\phi_{0} \equiv \hbar c/q$ and 
$\phi_{1}^{\scsc (j)}$ ($\phi_{2}^{\scsc (j)}$) is the integral 
of $A(x)$ over the upper (lower) leg between the $j$-th 
and the $j\!+\!1$-th rungs with $A(x)$ being the vector 
potential element in the direction of the legs at 
position $x$.  
   Here the matrix $t^{\scsc (j)}\equiv (t_{ll'}^{\scsc (j)})$ is 
the corresponding scattering matrix in the case that the vector 
potential in the legs is zero.  
   In this model the scattering matrix $T^{\scsc (j)}$ is  
the same as the corresponding transfer matrix, 
so the scattering matrix $\caS^{\scsc (j)} \equiv 
(\caS_{ll'}^{\scsc (j)})$ of the sub-system consisting of  
the first $j$ number of rungs is given by 
$T^{\scsc (j)} T^{\scsc (j-1)} \cdots 
T^{\scsc (1)}$. 
     
  We consider the injected current density 
$J_{\scsc (\mu,1)}^{\scsc (j)}$ ($J_{\scsc (\mu,2)}^{\scsc (j)}$) 
in the upper (lower) leg between the $j$-th and the 
$j\!+\!1$-th rungs, which is caused by the incident 
current $\bfa_{\mu}$ shown in Fig. \ref{ladde}.  
   This current is represented as $J_{\scsc (\mu,\nu)}^{\scsc (j)} 
= q \rho_{0} v |\caS_{\nu\mu}^{\scsc (j)}|^{2}$ where $v$ is 
the particle velocity in the upper (or lower) leg 
and $\rho_{0}$ is the local density of states $1/(2\pi\hbar |v|)$ 
in the one-dimensional perfect wire.

   Now we connect the scattering matrix $\caS^{\scsc (N)}$ of 
the system consisting of $N$ number of the rungs ($N\!>\!1$) to 
the injected current density $J_{\scsc (\mu,\nu)}^{\scsc (j)}$. 
   Noting unitarity of the matrices  
$T^{\scsc (j')}$, ${\scriptstyle j'=N,N-1,\cdots,l+1}$ and using 
the relation $\delta \phi_{\nu}^{\scsc (j)} / 
\delta A(x_{\nu'}^{\scsc (j')}) = \delta_{jj'}\delta_{\nu\nu'}$, 
where $x=x_{1}^{\scsc (j)}$ ($x_{2}^{\scsc (j)}$) is a point in 
the upper (lower) leg between the $j$-th and the 
$j\!+\!1$-th rungs, we obtain $(1/(2\pi i)) \sum_{l=1}^{2} 
\caS_{l\mu}^{\scsc (N) \dagger} \cdot \delta 
\caS_{l\mu}^{\scsc (N)} / \delta A(x_{\nu}^{\scsc (j)}) = 
(1/c) J_{\scsc (\mu,\nu)}^{\scsc (j)}$, which is just the injected  
current density formula (\ref{Curre}) in the time-independent 
Hamiltonian case.


\sect{Conclusion and remarks}
   In this Letter we have discussed formulas to derive 
the charge current density, the charge density and their quantum 
mechanical current density correlations from the scattering matrix 
in one-particle and time-dependent Hamiltonian systems.
   The gauge invariance requires Eq. (\ref{GaugeInvar3}) 
which has to be satisfied by these local quantities. 
   Using our formulas we obtained a relation between the local 
density of states and the total current density produced by 
incident particles at an energy in time-independent Hamiltonian 
systems. 
   Specifically we verified the current density formula for a chiral 
ladder model. 

   In this Letter for simplicity we considered 
one-particle systems only. 
   However our technique to derive Eqs. (\ref{Sumrule1}) and 
(\ref{Sumrule2}) can be used to obtain some generalizations to  
the formulas including many-particles' effects. 
   For instance, we can generalize Eq.  (\ref{Curre}) to the 
inelastic scattering cases by dynamical scatterers with no charge. 
   Eqs. (\ref{Sumrule1}) and (\ref{Sumrule2}) also suggest that 
we can derive formulas for other local physical quantities from 
functional derivatives of the scattering matrix with respect to 
other local fields. 
   For example, if the potential $U(\bfx,t)$ includes the term 
$-q\hat{\bfs}\cdot\bfB(\bfx,t)/2$ as an interaction effect of a 
spin $\hat{\bfs}$ with a magnetic field $\bfB(\bfx,t)$, then we 
can calculate the local expectation value of a spin component from 
a functional derivative of the scattering matrix with 
respect to the magnetic field.

   As another approach to the charge current density we can use 
the linear response theory \cite{kub57,bar89}. 
   The connection of the scattering theoretical approach 
to the linear response theoretical approach in the description 
of local quantities is left as a future problem.


\begin{center}$\ast\ast\ast$\end{center}

  
   I am very grateful to M. B\"uttiker for stimulating 
discussions, encouragements and a careful reading of this Letter. 
   Especially he gave me valuable comments and suggestions 
relating to the injectivity formula, time-dependent Hamiltonian problems 
and the gauge invariance, and introduced the chiral ladder model to me.



\end{multicols}
\end{document}